# Spontaneous emission modulation in biaxial hyperbolic van der Waals material

Haotuo Liu[1,2,3], Yang Hu[3,4,5], Qing Ai[1,2,*], Ming Xie[1,2], and Xiaohu Wu[3,**]

1 School of Energy Science and Engineering, Harbin Institute of Technology, Harbin 150001, P. R. China

2 Key Laboratory of Aerospace Thermophysics, Ministry of Industry and Information Technology, Harbin 150001, P. R. China

3 Shandong Institute of Advanced Technology, Jinan 250100, P. R. China

4 Basic Research Center, School of Power and Energy, Northwestern Polytechnical University, Xi'an, Shanxi 710072, P. R. China

5 Center of Computational Physics and Energy Science, Yangtze River Delta Research Institute of NPU, Northwestern Polytechnical University, Taicang, Jiangsu 215400, P. R. China

*Corresponding author: hitaiqing@hit.edu.cn (Q. Ai).

**Corresponding author: xiaohu.wu@iat.cn (X. Wu).

**Abstract:** As a natural van der Waals crystal, $\alpha$-MoO$_3$ has excellent in-plane hyperbolic properties and essential nanophotonics applications. However, its actively tunable properties are generally neglected. In this work, we achieved active modulation of spontaneous emission from a single-layer flat plate using the rotation method for the first time. Numerical results and theoretical analysis show that $\alpha$-MoO$_3$ exhibits good tunability when rotated in the *y-z* or *x-y* plane. A modulation factor of more than three orders of magnitude can be obtained at 634 cm$^{-1}$. However, when the rotation is in the *x-z* plane, the spontaneous emission of the material exhibits strong angle independence. The theoretical formulation and the physical mechanism analysis explain the above phenomenon well. In addition, for the semi-infinite $\alpha$-MoO$_3$ flat structure, we give the distribution of the modulation factor of spontaneous emission with wavenumber and rotation angle. Finally, we extended the calculation results from semi-infinite media to finite thickness films. We obtained the general evolution law of the peak angle of the modulation factor with thickness, increasing the modulation factor to about 2000. We believe that the results of this paper can guide the active modulation of spontaneous emission based on anisotropic materials.

## 1. Introduction

Spontaneous emission (SE) from quantum emitters is considered an intrinsic property of luminescent materials and is essential for micro-nano lasers, nanophotonics, and quantum computing [1-3]. According to Fermi's golden rule, it can be attributed to light-matter interactions and is proportional to the local density of optical states (LDOS) [4-8]. In general, the SE rate of quantum emitters can be tuned by changing the electromagnetic environment in which it is located. However, this method cannot achieve modulation of its SE rate when the electromagnetic environment is fixed. There has been evidence that active modulation of SE rate can be achieved by tuning the LDOS of the emitter, which has attracted significant research interest. Li et al. designed a metal-dielectric multilayer structure to achieve tunable LDOS enhancement by adjusting the metal filling ratio [9]. Hao et al. increased the SE rate by about two orders of magnitude using liquid crystals and low refractive index metamaterials [10]. T. Debu et al. proposed a multilayer structure based on graphene-ferroelectric material, which can achieve active tuning of SE by changing the chemical potential of graphene [11]. However, the current method of active modulation of SE involves complex nanofabrication techniques, and the limited range of SE regulation also hinders its application and development in related fields.

Recently, the natural biaxial hyperbolicity found in van der Waals (vdWs) crystals could open new avenues for mid-infrared nanophotonics due to the superior hyperbolic phonon polarization effect [12-16]. Compared to conventional artificial hyperbolic materials, natural hyperbolic materials avoid complex photolithography, high ohmic loss, and non-local response [17-20]. $\alpha$-MoO$_3$ is an emerging natural vdWs material with three orthogonal crystal axes, three Reststrahlen bands, and good in-plane hyperbolic properties. Moreover, the lifetime of its polariton was found to be about ten times longer than that of hBN [21]. Previous works on this material have been mainly physical and chemical analyses, and there are fewer studies on its tunability. Active tuning of the radiation transmission properties of the hyperbolic material can be achieved by rotation modulation, resulting in better system performance [22, 23]. However, this tuning method has never been applied to the SE tuning of quantum emitters.

In this work, we propose to tune the SE of quantum emitters by rotation modulation for the first time and perform theoretical calculations of its modulation performance. By combining the rotation matrix of $\alpha$-MoO$_3$, we calculate the dependence of its reflection coefficient on the rotation angle and rotation axis, give the corresponding Purcell factor, and further obtain the modulation factor versus wavenumber. The theoretical explanation for the specific SE modulation phenomenon is given when $\alpha$-MoO$_3$ rotates in different planes. In addition, we calculate the dependence of the Purcell factor of the quantum emitter on the rotation angle at a specific wavenumber and give the distribution of the modulation factor. Finally, we extend the calculation of semi-infinite medium to finite thickness films and obtain the angular evolution law corresponding to the peak of the modulation factor. This study can guide the design of active SE modulation in anisotropic materials.

## 2. Models and Methods

Here, we propose the rotation method, which is used to achieve active modulation of $\alpha$-MoO$_3$ single-layer structure. **Fig. 1** gives three different modulation modes to rotate $\alpha$-MoO$_3$ in the $y$-$z$, $x$-$z$ and $x$-$y$ planes. The green lines show the crystalline axis direction of $\alpha$-MoO$_3$. The three curved arrows represent the three possible rotation directions, and the rotation angles are $\gamma_x$, $\gamma_y$, and $\gamma_z$, respectively. The thickness of the single-layer structure is $t$. Moreover, we set the calculation plane as $x$-$z$ and the incident light as $p$-polarization.

As a natural vdWs material, $\alpha$-MoO$_3$ is essential for the future development of applications such as nano-imaging, bio-detection, and radiation energy control [24-26]. The dielectric function along each crystalline axis direction can be determined by a Lorentz model [27]:

$$\varepsilon_i = \varepsilon_{\infty,i}\left(1 + \frac{\omega_{LO,i}^2 - \omega_{TO,i}^2}{\omega_{TO,i}^2 - \omega^2 - j\omega\Gamma_i}\right), \qquad (1)$$

where $\varepsilon_i$, $\varepsilon_{\infty,i}$ are the dielectric constant components on the crystal axis and the high-frequency dielectric constant. $\omega_{LO,i}$, $\omega_{TO,i}$, and $\Gamma_i$ correspond to the longitudinal LO, the transverse TO optic phonon frequencies, and the damping constant. The specific parameters of the above variables can be found in Ref.[28]. To further illustrate the excellent properties of $\alpha$-MoO$_3$, we calculated the dielectric function along different crystalline axes, as shown in **Fig. 2**. The lattice of $\alpha$-MoO$_3$ consists of octahedral unit cells with Mo-O bonds of different lengths in the three main crystalline axes, which causes the structure produces rich phonon modes with infrared activity along different crystal orientations [29]. In addition, $\alpha$-MoO$_3$ has three Reststrahlen bands, 545-851 cm$^{-1}$ ([100]), 820-972 cm$^{-1}$ ([001]), and 958-1010 cm$^{-1}$ ([010]), where the element in parentheses represents the direction of the phonon modes.

When $\alpha$-MoO$_3$ is rotated in different planes, the dielectric function tensor will be converted to the following form [30]:

$$\boldsymbol{\varepsilon}_{i,\gamma} = \mathbf{T}_{i,\gamma} \begin{pmatrix} \varepsilon_x & 0 & 0 \\ 0 & \varepsilon_y & 0 \\ 0 & 0 & \varepsilon_z \end{pmatrix} \mathbf{T}_{i,\gamma}^{-1}, \tag{2}$$

where $\mathbf{T}_{i,\gamma}$ is the rotation matrix, corresponding to *y-z*, *x-z*, *x-y* planes as $\begin{pmatrix} 1 & 0 & 0 \\ 0 & \cos\gamma_x & -\sin\gamma_x \\ 0 & \sin\gamma_x & \cos\gamma_x \end{pmatrix}$,

$\begin{pmatrix} \cos\gamma_y & 0 & -\sin\gamma_y \\ 0 & 1 & 0 \\ \sin\gamma_y & 0 & \cos\gamma_y \end{pmatrix}$, and $\begin{pmatrix} \cos\gamma_z & -\sin\gamma_z & 0 \\ \sin\gamma_z & \cos\gamma_z & 0 \\ 0 & 0 & 1 \end{pmatrix}$, respectively.

The interaction of light with hyperbolic materials significantly affects the SE of quantum emitters in the infrared range [31]. For convenience, we use the dimensionless quantity Purcell factor to describe the SE of the quantum emitters. When the moment of the dipole on the flat plate is along the *z*-direction, the Purcell spectrum can be described as [32]:

$$\frac{\Gamma}{\Gamma_0} = 1 + \frac{3}{2k_0^3} \operatorname{Re}\left( \int_0^\infty \frac{r_{pp}(\omega, k_x) e^{2ik_z d} k_x^3 dk_x}{k_z} \right), \tag{3}$$

where $\Gamma$, $\Gamma_0$ are the SE rates of a dipole near the plate, and in free space. $k_x$ and $k_z$ represent the dimensionless wavevector components. *d* is the distance from the dipole to the flat plate, and the total LDOS enhancement is the integrand of Eq. 3 [33]. $r_{pp}(\omega,k_x)$ is the reflection coefficient for *p*-polarized light, which can be calculated using the 4 × 4 matrix [34, 35].

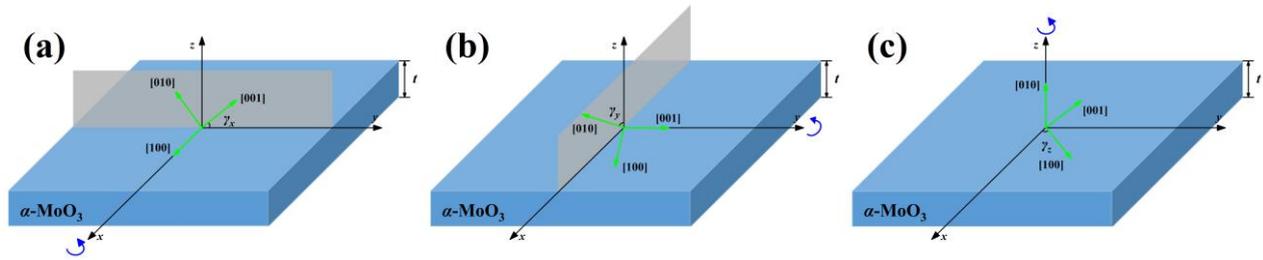

**Fig. 1** Schematic diagram of the rotation modulation in α-MoO₃ single-layer structure.

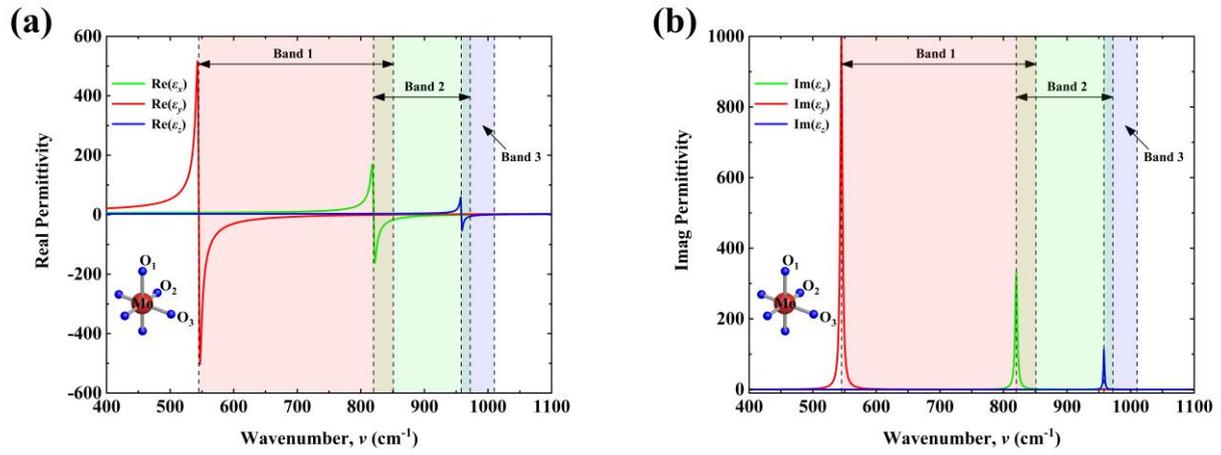

**Fig. 2** Dielectric functions along different crystal axes of α-MoO₃.

## 3. Results and Discussion

### 3.1 Rotation axis effects

To understand the modulation mechanism of SE, we first reduce the computational model to a semi-infinite medium. **Fig. 3** shows the Purcell factor for $α$-MoO$_3$ rotation in different planes. When $α$-MoO$_3$ is not rotated, the Purcell factor is significantly enhanced mainly in band 2 and 3. It is noteworthy that sharp peaks are found in the 958-972 cm$^{-1}$. Since the dielectric constants are less than zero in the *x*., *z*-direction and greater than zero in the *y*-direction at this time, the phenomenon can be attributed to the coupling of volume-confined hyperbolic polaritons (v-HPs) and surface-confined hyperbolic polaritons (s-HPs) [36]. When $γ_x$ = 90 deg, a redshift of the Purcell spectrum can be found accompanied by an increase in width. In addition, we can see that the Purcell spectra are roughly similar in shape, which can be explained by the excitation of band 1 caused by rotation, making 820-851 cm$^{-1}$ satisfying the coupled excitation conditions v-HPs and s-HPs. When $γ_y$ = 90 deg, the Purcell spectrum hardly changes, as explained in *Section 3.2*. Further, we calculated the case of $γ_z$ = 90 deg. Since this rotation avoids the overlap between bands, the Purcell spectrum becomes smooth and enhanced in both band 1 and 3. To compare the difference in SE modulation in different rotation planes, we define the modulation factor (ratio of Purcell factor in rotated case to that in unrotated case). As shown in **Fig. 3**, By making $α$-MoO$_3$ rotate 90 deg in the *x*-*y* plane, we can obtain a modulation factor of up to 400.

### 3.2 Mechanism analysis

To explain the mechanism of the effect of rotation on the active modulation of SE, we give the distribution of the reflection coefficient and the total LDOS enhancement in the wavevector space. **Fig. 4** (a) and (b) are the variations of the imaginary part of the reflection coefficient with wavenumber and dimensionless wavevector for $γ_x$ = 0/90 deg. The significance of the red dashed line remains the same as in **Fig. 2** and the black dashed line. As $α$-MoO$_3$ rotates in the *y*-*z* plane, the excitation region of the hyperbolic phonon polaritons shifts from band 3 to band 1, which is the main reason for the redshift of the Purcell spectrum. To further explain the abnormal Purcell spectral when $α$-MoO$_3$ is rotated in the $x$-$z$ plane, we give the formula for the reflection coefficient in this case [37]:

$$r = \frac{\sqrt{\varepsilon_x \varepsilon_z k_z^2} - i\sqrt{k_x^2 - \varepsilon_z \cos^2 \gamma_y - \varepsilon_x \sin^2 \gamma_y}}{\sqrt{\varepsilon_x \varepsilon_z k_z^2} + i\sqrt{k_x^2 - \varepsilon_z \cos^2 \gamma_y - \varepsilon_x \sin^2 \gamma_y}} . \tag{4}$$

When the dimensionless wavevector $k_x$ is large enough, $-\varepsilon_z \cos^2 \gamma_y - \varepsilon_x \sin^2 \gamma_y$ can be viewed as a negligible value. This indicates that the rotation of $α$-MoO$_3$ in the $x$-$z$ plane has no effect on the reflection coefficient at the large wavevector. **Fig. 4** (d) gives the deviation of reflection coefficient versus rotation angle $\gamma_y$ and dimensionless wavevector $k_\rho/k_0$. This indicates the drastic effect of rotating $α$-MoO$_3$ in the $x$-$z$ plane on the reflection coefficient when the wavevector is small. **Fig. 4** (e) compares the total LDOS enhancement for different rotation angles. The results show that the difference in reflection coefficients under the wavevector has almost no effect on the total LDOS enhancement, indirectly confirming that the Purcell factor is independent of the $\gamma_y$ in **Fig. 3**. Finally, we analyze the in-plane rotation and give the distribution of the imaginary part of the reflection coefficient along the $x$, $y$ direction dimensionless wavevector at 634 cm$^{-1}$ corresponding to the maximum modulation factor as shown in **Fig. 4** (c). We can conclude that the SE enhancement at this frequency is mainly attributed to v-HPs, and the excitation range is $-\sqrt{-\frac{\varepsilon_x}{\varepsilon_y}} < \frac{k_y}{k_x} < \sqrt{-\frac{\varepsilon_x}{\varepsilon_y}}$ [12].

At wavenumber of 634 cm$^{-1}$, the dielectric constant components of $α$-MoO$_3$ are $\varepsilon_x$ = 8.0283 + $j$0.0378 and $\varepsilon_y$ = -15.9569 + $j$0.5113. $\beta$ = 35.3 deg represents the angle of the v-HPs boundary with the horizontal line.

## *3.3 Rotation angle effects*

Further, we take the wavenumber of 634 cm$^{-1}$ corresponding to the maximum modulation factor in **Fig. 3**, and the effect of the rotation angle on the modulation factor is explored in-depth, as shown in **Fig. 5**. We can see that the modulation factor is almost constant when $α$-MoO$_3$ is rotated in the $x$-$z$ plane, which is consistent with the explanation in ***Section 3.2***. When $α$-MoO$_3$ is rotated in the $x$-$y$ plane, a peak modulation factor can be found at a specific angle with a maximum value of more than 1000. The peak angle is 37 deg, which is close to the $β$ in **Fig. 4** (c) in ***Section 3.2***, and further verifies the correctness of the physical mechanism explanation. When $α$-MoO$_3$ rotates in the $y$-$z$ plane, we can obtain a smaller peak angle of 24 deg. The excitation range of v-HPs that should be considered is $-\sqrt{-\frac{\varepsilon_z}{\varepsilon_y}} < \frac{k_y}{k_z} < \sqrt{-\frac{\varepsilon_z}{\varepsilon_y}}$. The dielectric constant components of $α$-MoO$_3$ are $\varepsilon_y$ = -15.9569 + $j$0.5113 and $\varepsilon_z$ = 2.8199 + $j$0.001. We can obtain $β$ = 22.8 deg close to the peak angle of 24 deg. The above results strongly demonstrate that the peak angle is almost the same as the angle between the boundary of the v-HPs excitation region and the horizontal line for semi-infinite media. The relationship between the modulation factor and wavenumber and rotation angle was further calculated because the modulation factor peaks exist at a specific angle when $α$-MoO$_3$ is rotated in different planes. **Fig. 6** exhibits regular enhancement and weakening regions (the shapes of the regions are approximately trapezoidal and triangular when the optical axes are rotated in the $y$-$z$ and $x$-$y$ planes, respectively). The enhancement region is mainly present at 545-820 cm$^{-1}$, where the maximum modulation factor is obtained near the beveled edge. The location of the weakened region is different, and it is located at 972-1010 cm$^{-1}$ and 820-958 cm$^{-1}$, respectively.

*3.4 Thickness effects*

The above-presented modulations of the SE of molybdenum trioxide plates are for semi-infinite media. Further extension of the scope of application for modulating SE by rotating $α$-MoO$_3$, we have performed calculations using finite thickness plates as shown in **Fig. 7**. It can be seen that the contribution to the modulation factor of the $t$ = 100 nm flat plate is almost the same as that of the semi-infinite flat plate. As the thickness gradually decreases, it can be found that the modulation factor also gradually increases. When the thickness is 1 nm, the modulation factor exceeds 2000. In addition, we can see that the peak angle also increases, which may be because v-HPs are more sensitive to the thickness, enhancing the SE when the plate thickness is minimal. The distribution of the imaginary part of the reflection coefficient in the dimensionless wavevector space for different thicknesses is given in **Fig. 8**. The red dashed line represents the dispersion relation of the v-HPs in $α$-$MoO_3$, which can be calculated using Eq. 71 from Ref.[38]. As the plate thickness decreases, the reflection coefficient gradually increases and moves towards the bigger wavevector direction, which directly causes the increase in the peak angle.

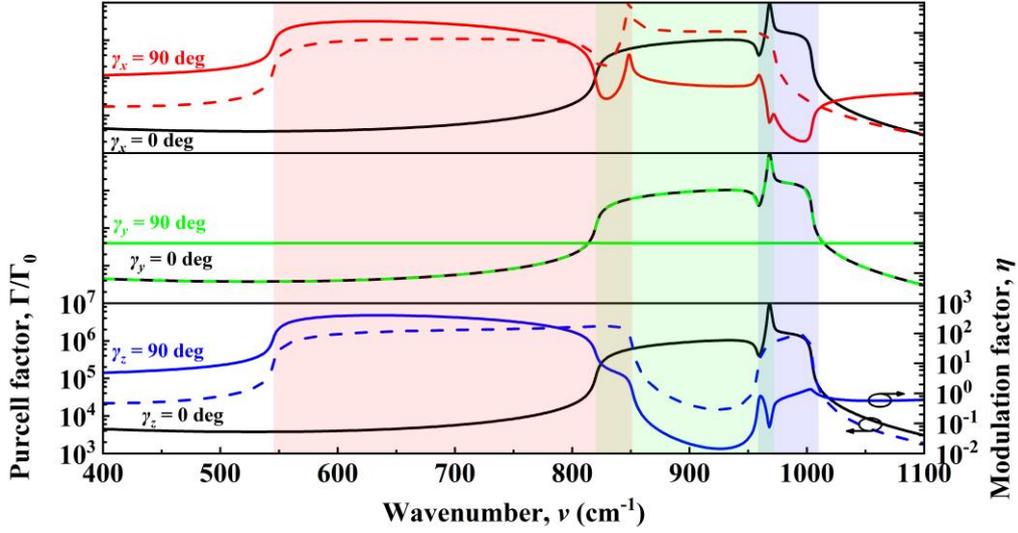

**Fig. 3** Effect of Purcell factor and modulation factor on wavenumber when *α*-MoO$_3$ is rotated in different planes.

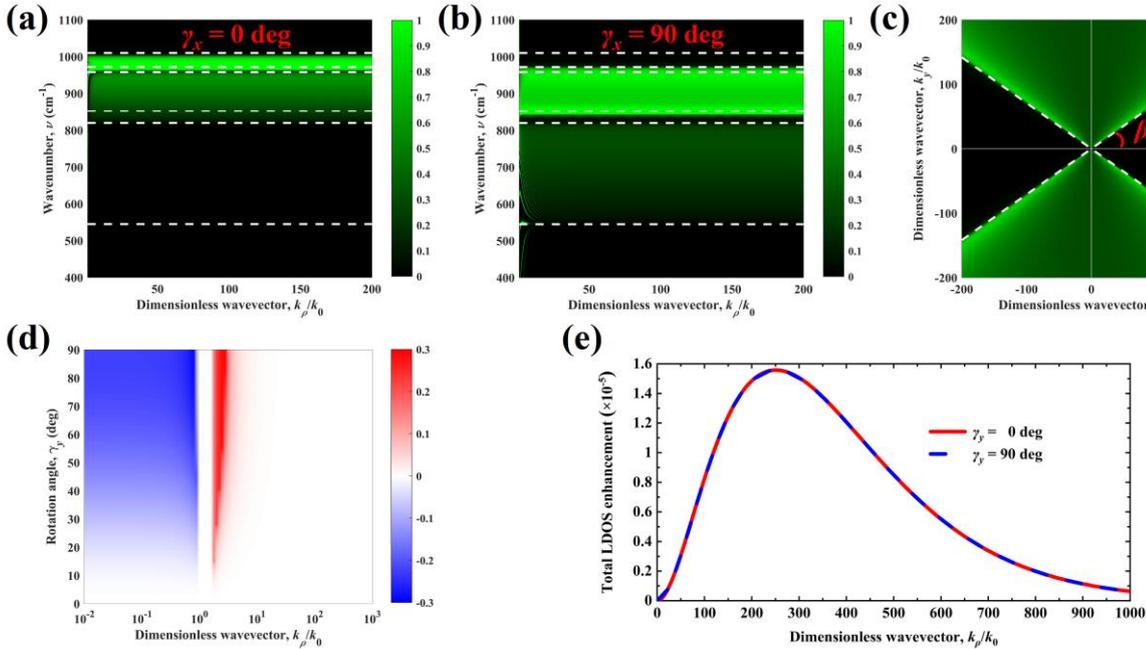

**Fig. 4** (a), (b) Dispersion curves of hyperbolic phonon polaritons in *α*-MoO$_3$ when $\gamma_x = 0/90$ deg. (c) The absolute value of the imaginary part of the reflection coefficient $r_{pp}$ varies with the dimensionless wavevector $k_x/k_0$ and $k_y/k_0$. (d) Contour of the distribution of the deviation of reflection coefficient versus rotation angle $\gamma_y$ and dimensionless wavevector $k_p/k_0$. (e) Total LDOS enhancement versus dimensionless wavevector $k_\rho/k_0$ at $\gamma_y = 0/90$ deg.

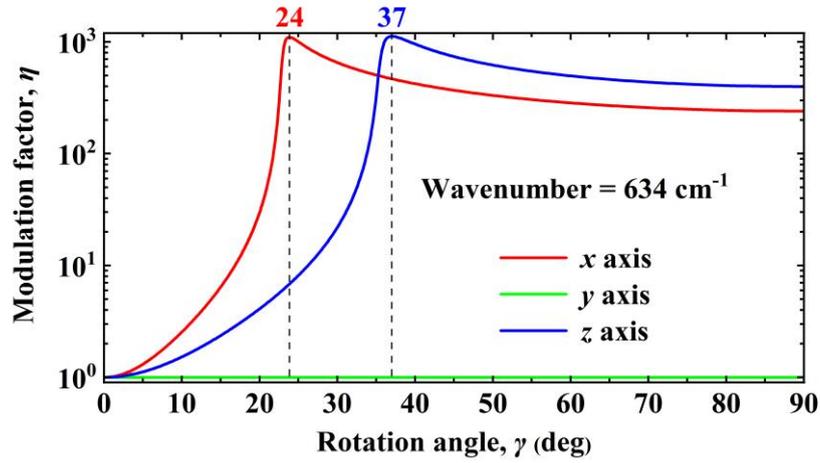

**Fig. 5** Modulation factor versus rotation angle $\gamma_x$, $\gamma_y$ and $\gamma_z$ when the wavenumber is 634 cm$^{-1}$.

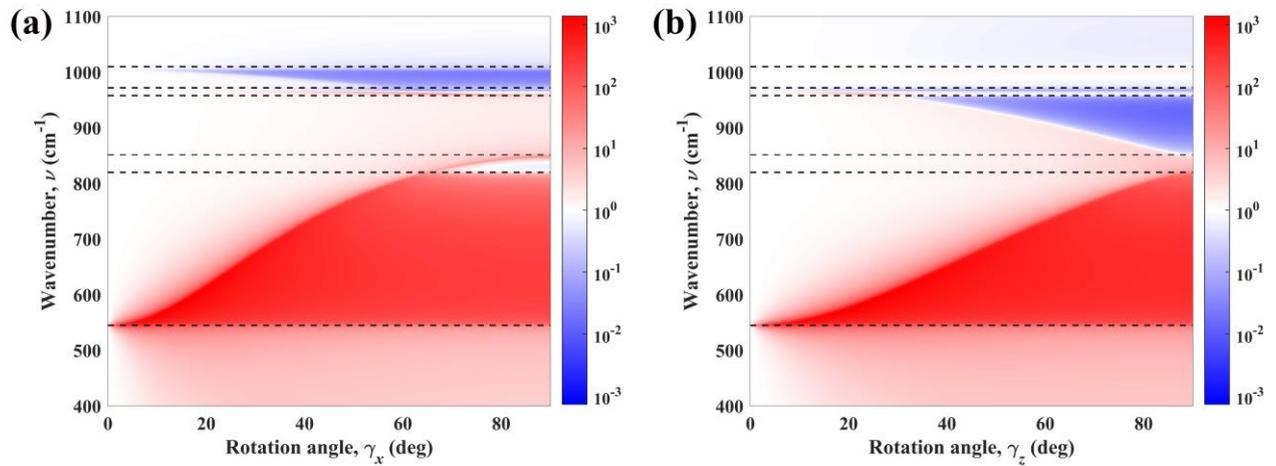

**Fig. 6** Modulation factor versus wavenumber and rotation angle $\gamma_x$, $\gamma_z$.

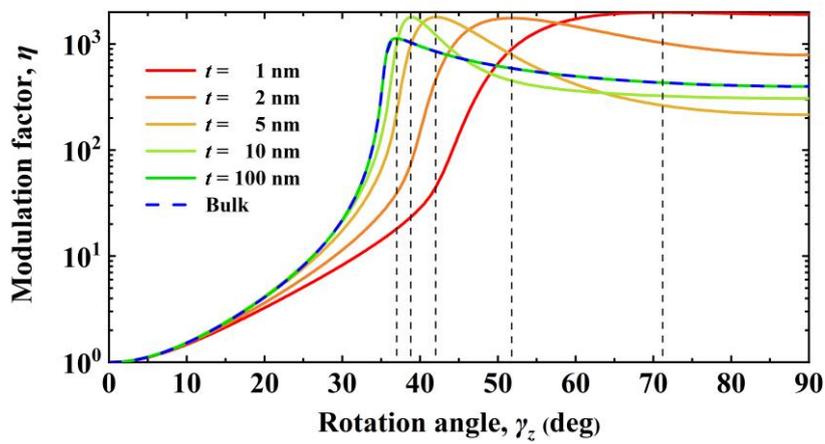

**Fig. 7** Modulation factor versus rotation angle $\gamma_z$ for different thicknesses.

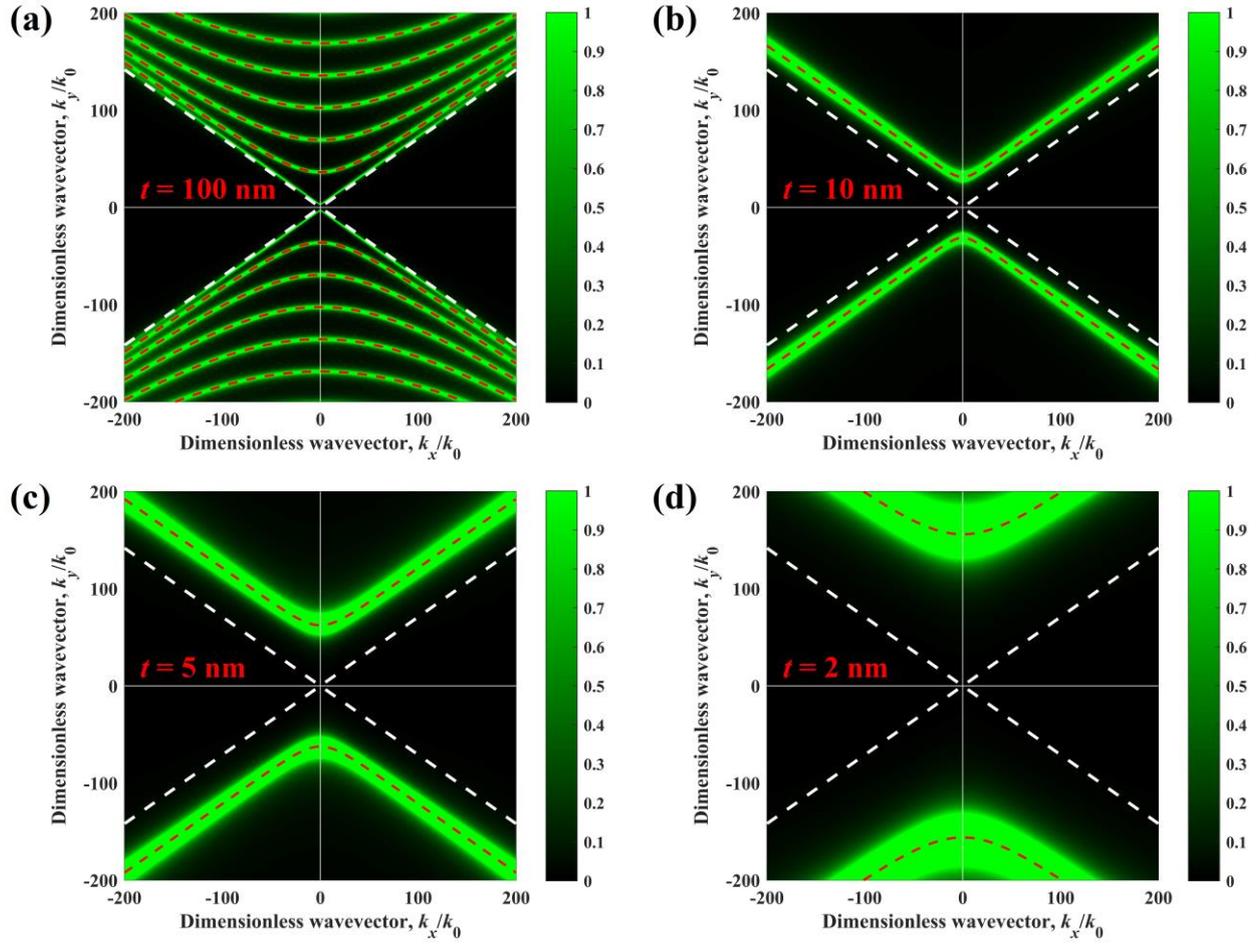

**Fig. 8** The absolute value of the imaginary part of the reflection coefficient $r_{pp}$ with wavevector components $k_x$ and $k_y$ at 634 cm$^{-1}$ for different values of the slab thickness. The white dashed line is $k_y = \pm 0.71 k_x$.

## 4. Conclusion

We proposed for the first time to use rotation to regulate the SE of a flat plate structure and numerically calculate its regulation performance. The results show that by rotating $α$-MoO$_3$ in the $y$-$z$ and $x$-$y$ planes, the modulation factor of the SE of a semi-infinite $α$-MoO$_3$ plate can be exceeded by more than 3 orders of magnitude. Conversely, when $α$-MoO$_3$ is rotated in the $x$-$z$ plane, the SE is almost a constant value. The above phenomenon is well explained by further mechanism analysis. In addition, we give the distribution of modulation factors depending on the wavenumber and the rotation angle. Finally, we calculated the modulation performance of a finite thickness flat plate. We found that decreasing the thickness can further enhance the modulation factor to about 2000 when the thickness is less than 100 nm. The reason can be explained as v-HPs are sensitive to thickness. This work will benefit the design of tunable quantum emitters based on vdWs materials.

## ACKNOWLEDGMENTS

This work was supported by the National Natural Science Foundation of China (NSFC) (51776052, 52106099), Aeronautical Science Foundation of China (ASFC) (201927077002), and Natural Science Foundation of Shandong Province (ZR2020LLZ004).